%% file: paper.tex
\documentclass[10pt,aps,prd,twocolumn,showpacs,amsmath,nofootinbib]{revtex4-1}

\usepackage[english]{babel}

\usepackage[caption=false]{subfig}
\usepackage{slashed}
\usepackage{dcolumn}
\usepackage{slashbox}

\usepackage{graphicx}
\usepackage{axodraw4j}

\usepackage[unicode=true]{hyperref}
\hypersetup{pdftitle=Radiative corrections to the Dalitz decay \textbackslash pi\^{}0 -> e\^{}+e\^{}-gamma revisited}
\hypersetup{pdfauthor=Tom\'{a}\v{s} Husek}

\newcommand*{\vv}[1]{\vec{\hspace{-0.3mm}\mkern2mu#1}\hspace{0.3mm}}

\begin{document}

\title{Radiative corrections to the Dalitz decay \texorpdfstring{$\pi^0\to e^+e^-\gamma$}{\textbackslash pi\^{}0 -> e\^{}+e\^{}-gamma} revisited}
\date{\today}
\author{Tom\'{a}\v{s} Husek}
\email{husek@ipnp.mff.cuni.cz}
\author{Karol Kampf}
\email{karol.kampf@mff.cuni.cz}
\author{Ji\v{r}\'{i} Novotn\'{y}}
\email{jiri.novotny@mff.cuni.cz}
\affiliation{Institute of Particle and Nuclear Physics, Faculty of Mathematics and Physics, Charles University in Prague, V Hole\v{s}ovi\v{c}k\'{a}ch 2, Praha 8, Czech Republic}

\begin{abstract}
We have recalculated the Mikaelian and Smith radiative corrections to the Dalitz decay $\pi^0\to e^+e^-\gamma$ beyond the soft-photon approximation, i.e.\ over the whole range of the Dalitz plot and with no restrictions on the radiative photon.
In contrast to the previous calculations, we did not neglect the terms of order higher than $\mathcal{O}(m^2)$ and also included the one-photon-irreducible contribution at one-loop level and the virtual muon loop contribution.
The results can then be used also for heavier particles in the final state.
\end{abstract}

\pacs{13.20.Cz, 13.40.Ks}

\maketitle


\section{Introduction}
\label{sec:intro}

Right after the process $\pi^0\to\gamma\gamma$, the second most important decay channel of a neutral pion is the Dalitz decay $\pi^0\to e^+e^-\gamma$ with a branching ratio $(1.174\pm0.035)\,\%$~\cite{Agashe:2014kda}.
This decay was named after Richard~H.~Dalitz, who first studied it in Ref.~\cite{Dalitz:1951aj}.
Experimental data of this process provide information about the semi-off-shell pion transition form factor $\mathcal{F}_{\pi^0\gamma\gamma^*}({Q^2}/{M^2})$ in 
the timelike region and in particular its slope parameter $a$.

Radiative corrections to the total decay rate of the Dalitz decay $\pi^0\to e^+e^-\gamma$ were first addressed by D.~Joseph~\cite{Joseph:1960zz}.
The pioneering study of the corrections to the differential decay rate was done by B.~E.~Lautrup and J.~Smith~\cite{Lautrup:1971ew} using the soft-photon approximation.
This analysis was soon after extended by K.~O.~Mikaelian and J.~Smith~\cite{Mikaelian:1972yg} by hard-photon corrections to the whole range of the bremsstrahlung photon energy.
As one of the main results of their work the table of radiative corrections $\delta(x,y)$ to the leading-order (LO) differential decay rate was presented.

It turned out that such a table would be very useful for the Monte Carlo simulations in experiments covering $\pi^0$ decays, e.g.\ the NA48  
experiment at CERN~\cite{Batley:2015lha}.
In practice, for the table of values $\delta(x,y)$, which was published in Ref.~\cite{Mikaelian:1972yg}, an interpolation or extrapolation procedure needs to be used in order to get the radiative correction at any desired point of the Dalitz plot.
This might lead to a large uncertainty.

We have therefore recalculated, generalized and extended the results presented in Ref.~\cite{Mikaelian:1972yg} and prepared the code which can give a value at any kinematical point $(x,y)$.
As we have not neglected the higher-order terms in the electron mass and included also the muon loop contribution to the vacuum polarization insertion correction, our result can be in principle also applied to the other related processes.
The decay of an eta meson to a muon pair and a photon, where the masses of the final-state particles are not anymore negligible in comparison to the decaying pseudoscalar, is such an example.
On the other hand, when an eta meson and its decays come into play, some peculiarities inevitably appear.
We comment on this a little in the present work but postpone the details and the results of the radiative corrections for this case to the paper in preparation.
Nevertheless, we try to be as general as possible considering the presented results so one can utilize the formulas without modifications later on.

To proceed even further we have also included the one-loop one-photon-irreducible contribution, which was considered to be negligible in the original paper~\cite{Mikaelian:1972yg} due to its proportionality to the lepton mass.
This statement had been corrected in Ref.~\cite{Tupper:1983uw} many years before the debate about this issue was closed; see e.g.\ Refs. \cite{Lambin:1985sb,Tupper:1986yk}.
We provide here a complete calculation of this contribution making no approximations considering the lepton masses and energy of the photon.
We show that this correction is indeed important and changes significantly the values of entries stated in Table I of Ref.~\cite{Mikaelian:1972yg} especially for a large invariant dilepton mass.

Let us also mention that a systematic treatment of the next-to-leading-order (NLO) corrections to the Dalitz decay of a neutral pion in the framework of chiral perturbation theory with dynamical leptons and photons was studied in Ref.~\cite{Kampf:2005tz}.
Here we will also use some results of this work.

It is worth it to notice that throughout the paper we stick to the notation which was used in Ref.~\cite{Mikaelian:1972yg} using only minor modifications.
Even though some of the names may appear to be clumsy, we believe that it would be confusing to do otherwise.
Naturally, such an approach is also very convenient for the reader who is familiar with the original work.

Our paper is organized as follows.
We recapitulate first some basic facts about the LO differential decay width calculation in Sec.~\ref{sec:LO}. 
Then we proceed to the review of the NLO radiative corrections in the QED sector in Secs.~\ref{sec:virt}, \ref{sec:1gI} and \ref{sec:BS}.
In particular, in Sec.~\ref{sec:virt} we discuss the virtual corrections including the muon loop contribution, in Sec.~\ref{sec:1gI} we introduce the one-photon-irreducible contribution and in Sec.~\ref{sec:BS} we describe the bremsstrahlung correction calculation.
Some technical details together with extensive results concerning the bremsstrahlung contribution to the NLO correction have been moved to the Appendixes.


\section{Leading order}
\label{sec:LO}

\begin{figure}[t!]
\centering
\setlength{\unitlength}{0.6pt}
  \begin{picture}(255,145) (139,-190)
    \SetScale{0.6}
    \SetWidth{1.0}
    \SetColor{Black}
    \Line[dash,dashsize=10](160,-123)(230,-123)
    \Photon(306,-177)(234,-123){4}{5.5}
    \Line[arrow,arrowpos=0.5,arrowlength=5,arrowwidth=2,arrowinset=0.2](342,-123)(288,-87)
    \Line[arrow,arrowpos=0.5,arrowlength=5,arrowwidth=2,arrowinset=0.2](288,-87)(342,-51)
    \Photon(234,-123)(288,-87){4}{4.5}
    \GOval(240,-123)(10,10)(0){0.882}
  \end{picture}
\caption{
\label{fig:LO}
Leading order diagram of the Dalitz decay $\pi^0\to e^+e^-\gamma$ in the QED expansion.
}
\end{figure}
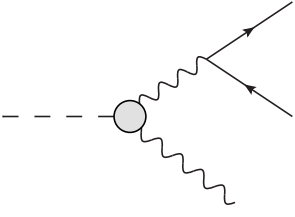

First, let us briefly introduce some basic notation.
In what follows we denote the four-momenta of the neutral pion (of the mass $M$), electron (mass $m$), positron and photon by $P$, $p$, $q$ and $k$, respectively.
We also introduce common kinematic variables $x$ and $y$ defined as
\begin{equation}
x=\frac{(p+q)^2}{M^2}\,,\quad y=-\frac{2}{M^2}\frac{P\cdot(p-q)}{(1-x)}\,,
\label{eq:defxy}
\end{equation}
where $x$ is a normalized square of the total energy of the $e^+e^-$ pair in their center-of-mass system (CMS), or simply of the electron-positron pair invariant mass. The variable $y$ has then the meaning of the rescaled cosine of the angle between the directions of the outgoing photon and positron in the $e^+e^-$ CMS. If we introduce $\nu=2m/M$ and
\begin{equation}
\beta=\beta(x)=\sqrt{1-\frac{\nu^2}{x}}\,,
\end{equation}
we can write the limits on $x$ and $y$ as
\begin{equation}
x\in [\nu^2,1]\,,\quad
y\in [-\beta,\beta]\,.
\end{equation}
The leading-order diagram of the Dalitz decay $\pi^0\to e^+e^-\gamma$ is shown in Fig.~\ref{fig:LO}.
The shaded blob corresponds to the neutral pion semi-off-shell transition form factor%
\footnote{As it also follows from the definition (\ref{eq:TFF}), we will use shortly $\mathcal{F}(0)=\mathcal{F}_{\pi^0\gamma\gamma^*}(0)$, which is complementary to the doubly off-shell transition form factor taken at the photon point $\mathcal{F}_{\pi^0\gamma\gamma^*}(0)=\mathcal{F}_{\pi^0\gamma^*\gamma^*}(0,0)\equiv\mathcal{F}_{\pi^0\gamma\gamma}$, which at the LO of the chiral expansion is equal to $\mathcal{F}_{\pi^0\gamma\gamma}^\text{LO}=-1/(4\pi^2F)$.}
\begin{equation}
\mathcal{F}\bigg(\frac{Q^2}{M^2}\bigg)
=\mathcal{F}_{\pi^0\gamma\gamma^*}\bigg(\frac{Q^2}{M^2}\bigg)
\equiv\mathcal{F}_{\pi^0\gamma\gamma^*}(0)\,f\bigg(\frac{Q^2}{M^2}\bigg)\,,
\label{eq:TFF}
\end{equation}
which is related to the doubly off-shell transition form factor $\mathcal{F}_{\pi^0\gamma^*\gamma^*}(Q_1^2/M^2,Q_2^2/M^2)=\mathcal{F}_{\pi^0\gamma^*\gamma^*}(Q_2^2/M^2,Q_1^2/M^2)$, defined as
\begin{equation}
\begin{split}
&\int\text{d}^4x\,e^{il\cdot x}\langle0|T[j^\mu(x)j^\nu(0)]|\pi^0(P)\rangle\\
&=-i\epsilon^{\mu\nu\alpha\beta}l_\alpha P_\beta\mathcal{F}_{\pi^0\gamma^*\gamma^*}\big(l^2/M^2,(P-l)^2/M^2\big)\,,
\end{split}
\end{equation}
by $\mathcal{F}_{\pi^0\gamma\gamma^*}(Q^2/M^2)=\mathcal{F}_{\pi^0\gamma^*\gamma^*}(0,Q^2/M^2)$.
In Eq.~(\ref{eq:TFF}), $f$ is a dimensionless function, which can be linearly expanded in the chiral perturbation theory in terms of the slope parameter $a$ as follows
\begin{equation}
f(z)\simeq1+az\,.
\label{eq:fx}
\end{equation}
In our case it then holds $Q^2=(P-k)^2=M^2x$\, and for the leading-order matrix element in the QED expansion we can write
\begin{equation}
\begin{split}
i&\mathcal{M}^\text{LO}(p,q,k)
=\frac{e^3}{M^2x}\mathcal{F}(x)\epsilon^{*\rho}(k)\\
&\times\Big\{\,2m\left[\bar{u}(p,m)\gamma_\rho\slashed k\gamma_5 v(q,m)\right]\\
&+\Big[\bar{u}(p,m)\left[\gamma_\rho\left(k\cdot p\right)-p_\rho\slashed k\right]\gamma_5 v(q,m)\Big]\\
&-\Big[\bar{u}(p,m)\left[\gamma_\rho\left(k\cdot q\right)-q_\rho\slashed k\right]\gamma_5 v(q,m)\Big]\Big\}\,.
\end{split}
\label{eq:MD}
\end{equation}
Summing the modulus squared of the previous result over the fermion spins and photon polarizations and taking into account that, in general, in terms of variables $x$ and $y$ it holds
\begin{equation}
\text{d}\Gamma(x,y)
=\frac{M}{(8\pi)^3}\overline{\left|\mathcal{M}(x,y)\right|^2}(1-x)\,\text{d}x\,\text{d}y\,,
\label{eq:dGammaxy}
\end{equation}
the differential decay rate then reads
\begin{equation}
\begin{split}
&\frac{{\mathrm d}^2\Gamma^\text{LO}}{{\mathrm d} x{\mathrm d} y}
=\frac{M}{(8\pi)^3}\frac{e^6M^2}2|\mathcal{F}(x)|^2\\
&\hspace{11mm}\times\frac{(1-x)^3}{x}\left[1+y^2+\frac{\nu^2}{x}\right]\\
&=\left(\frac{\alpha}{\pi}\right)|f(x)|^2\,\Gamma_{\pi^0\to\gamma\gamma}^\text{LO}\frac{(1-x)^3}{4x}\left[1+y^2+\frac{\nu^2}{x}\right]\,.
\end{split}
\label{eq:dLOxy}
\end{equation}
Here we have used the LO expression for the decay rate of the neutral pion main decay mode 
\begin{equation}
\Gamma_{\pi^0\to\gamma\gamma}^\text{LO}
=\frac{e^4M^3}{64\pi}|\mathcal{F}(0)|^2\,.
\end{equation}
Integrating (\ref{eq:dLOxy}) over $y$ we find
\begin{equation}
\begin{split}
\frac{{\mathrm d}\Gamma^\text{LO}}{{\mathrm d} x}
&=\left(\frac{\alpha}{\pi}\right)|f(x)|^2\,\Gamma_{\pi^0\to\gamma\gamma}^\text{LO}\frac{8\beta}{3}\frac{(1-x)^3}{4x}\bigg[1+\frac{\nu^2}{2x}\bigg]\,.
\end{split}
\end{equation}

Moving beyond the leading order, it is convenient to introduce the NLO correction $\delta$ to the LO differential decay width, which can be in general defined as (in the case of the two-fold differential decay width)
\begin{equation}
\delta(x,y)
=\frac{{\mathrm d}^2\Gamma^\text{NLO}}{{\mathrm d} x{\mathrm d} y}\bigg/\frac{{\mathrm d}^2\Gamma^\text{LO}}{{\mathrm d} x{\mathrm d} y}
\label{eq:dxy}
\end{equation}
or (in the one-fold differential case)
\begin{equation}
\delta(x)
=\frac{{\mathrm d}\Gamma^\text{NLO}}{{\mathrm d} x}\bigg/\frac{{\mathrm d}\Gamma^\text{LO}}{{\mathrm d} x}\,.
\end{equation}
Such a correction can be divided into three parts emphasizing its origin
\begin{equation}
\delta
=\delta^\text{virt}+\delta^{1\gamma\text{IR}}+\delta^\text{BS}\,.
\end{equation}
Here, $\delta^\text{virt}$ stands for the virtual radiative corrections, $\delta^{1\gamma\text{IR}}$ for the one-photon-irreducible contribution, which is treated separately from $\delta^\text{virt}$ in our approach, and $\delta^\text{BS}$ for the bremsstrahlung.
Having knowledge of $\delta(x,y)$, we can calculate $\delta(x)$ as a trivial consequence of previous equations using the prescription
\begin{equation}
\delta(x)=
\frac 3{8\beta}\frac1{(1+\frac{\nu^2}{2x})}\int_{-\beta}^\beta\delta(x,y)\left[1+y^2+\frac{\nu^2}{x}\right]{\mathrm d} y\,.
\label{eq:dx}
\end{equation}
In the following sections, we discuss the individual contributions one by one.


\section{Virtual radiative corrections}
\label{sec:virt}

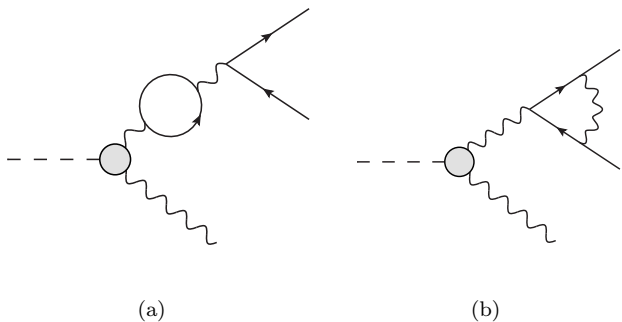
\begin{figure}[!ht]
\centering
\subfloat[][]{
\centering
\setlength{\unitlength}{0.58pt}
\begin{picture}(0,180) (260,-180)
    \SetScale{0.58}
    \SetWidth{1.0}
    \SetColor{Black}
    \Line[dash,dashsize=10](164,-96)(224,-96)
    \Photon(300,-150)(228,-96){4}{5.5}
    \Line[arrow,arrowpos=0.5,arrowlength=5,arrowwidth=2,arrowinset=0.2](360,-70)(306,-34)
    \Line[arrow,arrowpos=0.5,arrowlength=5,arrowwidth=2,arrowinset=0.2](306,-34)(360,2)
    \Photon(288,-52)(306,-34){4}{1.5}
    \Photon(228,-96)(253,-71){4}{2}
    \Arc[arrow,arrowpos=0.5,arrowlength=5,arrowwidth=2,arrowinset=0.2](270,-61)(20.125,153,513)
    \GOval(234,-96)(10,10)(0){0.882}
\end{picture}
\label{fig:Va}
}
\hspace{4cm}
\subfloat[][]{
\centering
\setlength{\unitlength}{0.55pt}
\begin{picture}(0,180) (260,-210)
    \SetScale{0.55}
    \SetWidth{1.0}
    \SetColor{Black}
    \Line[dash,dashsize=10](170,-123)(230,-123)
    \Photon(306,-177)(234,-123){4}{5.5}
    \Line[arrow,arrowpos=0.65,arrowlength=5,arrowwidth=2,arrowinset=0.2](350,-128)(288,-87)
    \Line[arrow,arrowpos=0.35,arrowlength=5,arrowwidth=2,arrowinset=0.2](288,-87)(350,-46)
    \Photon(234,-123)(288,-87){4}{4.65}
    \PhotonArc(309,-87)(26,-60,60){3}{4.25}
    \GOval(240,-123)(10,10)(0){0.882}
\end{picture}
\label{fig:Vb}
}
\caption{
\label{fig:loop}
Virtual radiative corrections for $\pi^0\to e^+e^-\gamma$ process: vacuum polarization insertion (a) and correction to the QED vertex (b).
}
\end{figure}
From the interference terms of the LO diagram shown in Fig.~\ref{fig:LO} with the one-loop diagrams presented in Fig.~\ref{fig:loop}, we get NLO virtual radiative corrections which can be written as~\cite{Mikaelian:1972yg}
\begin{equation}
\delta^\text{virt}(x,y)\\
=2\operatorname{Re}\left\{-\Pi(x)+F_1(x)+\frac{2F_2(x)}{1+y^2+\frac{\nu^2}{x}}\right\}
\label{eq:dvirt}
\end{equation}
or (through the formula (\ref{eq:dx})) as
\begin{equation}
\delta^\text{virt}(x)\\
=2\operatorname{Re}\left\{-\Pi(x)+F_1(x)+\frac{3F_2(x)}{2\left(1+\frac{\nu^2}{2x}\right)}\right\}\,.
\end{equation}
For the correction stemming from the vacuum polarization insertion in Fig.~\ref{fig:loop}(a) we can write
\begin{equation}
\Pi(x)=\Pi_e(x)+\Pi_\mu(x)\,.
\label{eq:VPI}
\end{equation}
Here we have explicitly written not only the contribution coming from the electron loop as it was done in Ref.~\cite{Mikaelian:1972yg}, but also from the muon loop.
This becomes both necessary and convenient when one goes beyond the decay $\pi^0\to e^+ e^-\gamma$, which we discuss throughout this text, and proceeds to the process $\eta\to \mu^+\mu^-\gamma$.
In the end, we might then simply do the exchange $m_e\leftrightarrow m_\mu$ ($m_e$ and $m_\mu$ stand for the electron and muon mass, respectively) in the expression for the correction $\delta$ to vary the final-state lepton masses.
Let us remark that independently of the considered processes, the loop with the lightest fermion is of the greatest importance.
Thus, taking only the electron loop into account (i.e.\ leaving the muon part in (\ref{eq:VPI})) and performing simply the tempting lepton mass substitution in the whole expression, we would miss out a very important contribution.
Obviously, the vacuum polarization insertion defined in a way shown in (\ref{eq:VPI}) stays after such an operation intact, as desired.
The other option would be to treat separately the final-state lepton masses $m$ and the masses of the particles in the vacuum polarization insertion loops $m_e$ and $m_\mu$.
This more universal approach was used in the code which comes with the paper.
Let us now introduce for the later convenience
\begin{equation}
\gamma=\gamma(x)=\frac{1-\beta(x)}{1+\beta(x)}\,.
\end{equation}
The individual terms used in (\ref{eq:VPI}) are then defined as
\begin{equation}
\Pi_\ell(x)=\frac{\alpha}{\pi}\left[-\frac19+\frac13\left(1+\frac{\nu_\ell^2}{2x}\right)\Big(2+\beta_\ell\log[-\gamma_\ell]\Big)\right]\,.
\label{eq:VPIl}
\end{equation}
In the above formula, $\ell$ stands for $e$ or $\mu$ in the loop and changes the meaning of the so far used electron mass $m$ in the definitions of $\nu$, $\beta$ and $\gamma$ to $m_e$ or $m_\mu$.
Unlike in Ref.~\cite{Mikaelian:1972yg} where only the real part of (\ref{eq:VPIl}) above the threshold $x=\nu_\ell^2$ is shown, we quote here the full expression valid in all kinematical regimes.
This is necessary to get right the contribution from the charged fermion loop when the transferred momentum is not sufficiently large to produce the real pair, i.e. for $x<\nu_\ell^2$, and lacks therefore the imaginary part.
This situation for instance appears (at least for a part of the kinematical region) when the pseudoscalar decays to the electron-positron pair via the muon loop.
For the purpose of real algebra used in the code (i.e.\ to avoid complex logarithms and so on) we can extract the real part of (\ref{eq:VPIl}).
For an arbitrary mass of the charged loop fermion we find
\begin{equation}
\begin{split}
&\operatorname{Re}\big\{\beta_\ell\log[-\gamma_\ell]\big\}\\
&=-2|\beta_\ell|\left\{\theta(\beta_\ell^2)\operatorname{arctanh}\beta_\ell+\theta(-\beta_\ell^2)\operatorname{arctan}\frac1{|\beta_\ell|}\right\}.
\end{split}
\end{equation}
In the following, we stick exclusively back to the process $\pi^0\to e^+ e^-\gamma$ and $m$ then denotes the outgoing electron mass as before.
Finally, for the electromagnetic form factors $F_1(x)$ and $F_2(x)$ stemming from the QED vertex correction in Fig.~\ref{fig:loop}(b) we have
\begin{equation}
\begin{split}
F_1(x)
&=\frac{\alpha}{\pi}
\left\{
-1-\frac{1+2\beta^2}{4\beta}\log(-\gamma)\right.\\
&-\frac{1+\beta^2}{2\beta}\bigg[\text{Li}_2(1-\gamma)+\frac14\log^2(-\gamma)\\
&\hspace{18mm}-\frac{\pi^2}{4}-i\pi\log(1-\gamma)\bigg]\\
&\left.+\left[1+\frac{1+\beta^2}{2\beta}\log(-\gamma)\right]\log\frac m{\lambda}
\right\}
\end{split}
\label{eq:F1}
\end{equation}
and
\begin{equation}
F_2(x)
=\frac{\alpha}{\pi}\frac{\nu^2}{4x\beta}\log{(-\gamma)}\,.
\label{eq:F2}
\end{equation}
In the above formulas, $\text{Li}_2$ stands for the dilogarithm and $\lambda$ is the infrared cutoff.
To extract the real parts from the previous terms (\ref{eq:F1}) and (\ref{eq:F2}) (in a sense of applying the operator $\operatorname{Re}$), in the kinematically allowed region where $M^2x\ge4m^2$ we use $\log(-\gamma)=\log(\gamma)+i\pi$, since $0\le\gamma\le1$.
Thus it is straightforward to see that the real parts of $F_1(x)$ and $F_2(x)$ indeed coincide with the form factors stated in Ref.~\cite{Mikaelian:1972yg} including the Coulomb term proportional to $-\pi^2/2$.


\section{One-photon-irreducible virtual radiative correction}
\label{sec:1gI}

\begin{figure}[!ht]
\centering
\subfloat[][]{
\centering
\setlength{\unitlength}{0.5pt}
\begin{picture}(160,220) (190,-180)
    \SetScale{0.5}
    \SetWidth{1.0}
    \SetColor{Black}
    \Line[dash,dashsize=10](170,-53)(234,-53)
    \Photon(234,-53)(306,1){4}{5.5}
    \Line[arrow,arrowpos=0.5,arrowlength=5,arrowwidth=2,arrowinset=0.2](360,-143)(306,-107)
    \Line[arrow,arrowpos=0.5,arrowlength=5,arrowwidth=2,arrowinset=0.2](306,-107)(306,1)
    \Line[arrow,arrowpos=0.5,arrowlength=5,arrowwidth=2,arrowinset=0.2](342,25)(378,49)
    \Photon(306,-107)(234,-53){4}{5.5}
    \Photon(342,25)(378,1){4}{3.5}
    \Line[arrow,arrowpos=0.5,arrowlength=5,arrowwidth=2,arrowinset=0.2](306,1)(342,25)
    \Text(195,-170)[lb]{\large{\Black{+ cross + CT}}}
\end{picture}
\label{fig:T2}
}
\hspace{1cm}
\subfloat[][]{
\centering
\setlength{\unitlength}{0.5pt}
\begin{picture}(160,220) (190,-180)
    \SetScale{0.5}
    \SetWidth{1.0}
    \SetColor{Black}
    \Line[dash,dashsize=10](170,-88)(234,-88)
    \Photon(234,-88)(306,-34){4}{5.5}
    \Line[arrow,arrowpos=0.5,arrowlength=5,arrowwidth=2,arrowinset=0.2](360,-178)(306,-142)
    \Line[arrow,arrowpos=0.5,arrowlength=5,arrowwidth=2,arrowinset=0.2](306,-88)(306,-34)
    \Line[arrow,arrowpos=0.5,arrowlength=5,arrowwidth=2,arrowinset=0.2](306,-34)(360,2)
    \Photon(306,-142)(234,-88){4}{5.5}
    \Line[arrow,arrowpos=0.5,arrowlength=5,arrowwidth=2,arrowinset=0.2](306,-142)(306,-88)
    \Photon(306,-88)(366,-88){4}{4}
\end{picture}
\label{fig:T4}
}
\caption{
\label{fig:1gI}
One-loop one-photon-irreducible contribution LO Feynman diagrams for $\protect\pi^0\to e^+e^-\gamma$ process considering the QED and $\chi$PT expansion:
triangle diagrams and related counterterms (a) and box diagram (b).
Note that ``cross'' accounts for a diagram with a photon coming from the outgoing positron line.
``CT'' then stands for two counterterm diagrams necessary to compensate the UV divergent parts of the related Feynman diagrams.
}
\end{figure}
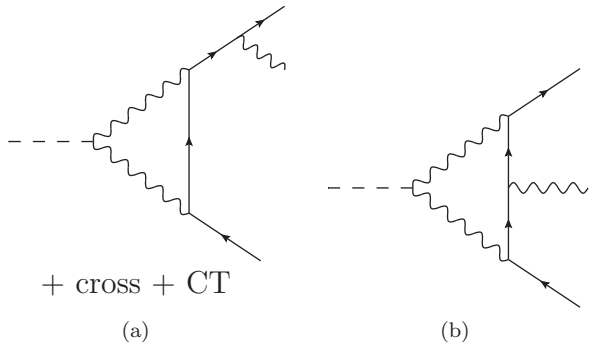

One-photon-irreducible (1$\gamma$IR) contributions were extensively studied in Ref.~\cite{Husek:2014tna} in connection with the bremsstrahlung correction to the $\pi^0\to e^+e^-$ process.
Here we will summarize the most important results which are necessary to proceed toward our purpose considering NLO corrections to the decay $\pi^0\to e^+e^-\gamma$ě.
Let us emphasize that this contribution was not included in the calculations performed in Ref.~\cite{Mikaelian:1972yg}.
On the other hand, it was shown later on in Ref.~\cite{Tupper:1983uw} within the limit $m\to0$ that there is no point in treating the 1$\gamma$IR correction as negligible.
In the following we show the results of the calculation beyond this massless limit.
For the reasons specified in the previous sentences we have devoted to this contribution a separate section, even though it is of course just one additional virtual radiative correction.

Until now we have not considered any particular form of the semi-off-shell form factor $\mathcal{F}(x)$ in our calculations.
To get the one-photon-irreducible contribution in a closed form, it is though necessary to choose a concrete form of $\mathcal{F}_{\pi^0\gamma^*\gamma^*}$.
Accordingly, we should consider at this moment a general doubly off-shell pion transition form factor $\mathcal{F}_{\pi^0\gamma^*\gamma^*}\big(l^2/M^2,(P-l)^2/M^2\big)$, where $l$ stands for a loop momentum.
In Fig.~\ref{fig:1gI}, we can see the LO of the considered contribution in chiral perturbation theory.
In such a limit we take the constant $\mathcal{F}_{\pi^0\gamma\gamma}^\text{LO}=-1/(4\pi^2F)$ as the local form factor and it is thus clear due to the power counting that a counterterm is needed.
The finite part of such a counterterm renormalized at scale $\mu$ is governed by the parameter $\chi^\text{(r)}(\mu)$, which corresponds to the high-energetic behavior of the complete form factor.
This can be theoretically modeled e.g.\ by the lowest-meson-dominance (LMD) approximation to the large-$N_C$ spectrum of vector-meson resonances yielding the value $\chi^\text{(r)}(M_\rho)=2.2\pm0.9$ \cite{Knecht:1999gb}, which can be further used for numerical results.
The dependance of the correction $\delta^{1\gamma\text{IR}}$ on $\chi^\text{(r)}$ can be neglected for the values given by relevant models as well as experiments, when a decay with the electrons in the final state is taken into account.
We will comment on this in the end of this section.
Let us emphasize in a more straightforward way that using ``only'' the LO expansion of the form factor is compensated by the effective value $\chi^\text{(r)}$ which differs for particular models.
One gets the model corresponding value of $\chi^\text{(r)}$ for instance from the matching to the full calculation.
In this sense one loses no information.
On the contrary, the model dependence of any such result can be conveniently altered easily just by changing the value of $\chi^\text{(r)}$.

The total matrix element covering all the diagrams represented in Fig.~\ref{fig:1gI} can be written in such a form which manifestly satisfies the Ward identities for the conserved electromagnetic vector current
\begin{equation}
\begin{split}
&i\mathcal{M}_{1\gamma\text{IR}}(p,q,k)=-\frac{i e^5}{2}\mathcal{F}_{\pi^0\gamma\gamma}^\text{LO}\epsilon^{*\rho}(k)\\
&\hspace{-2mm}\times\Big\{P(x,y)\left[\left(k\cdot p\right) q_\rho-\left(k\cdot q\right) p_\rho\right]\left[\bar{u}(p,m)\gamma_5 v(q,m)\right]\\
&+A(x, y)\Big[\bar{u}(p,m)\left[\gamma_\rho\left(k\cdot p\right)-p_\rho\slashed k\right]\gamma_5 v(q,m)\Big]\\
&-A(x,-y)\Big[\bar{u}(p,m)\left[\gamma_\rho\left(k\cdot q\right)-q_\rho\slashed k\right]\gamma_5 v(q,m)\Big]\\
&+T(x,y)\left[\bar{u}(p,m)\gamma_\rho\slashed k\gamma_5 v(q,m)\right]\Big\}\,.
\end{split}
\label{eq:M1gI}
\end{equation}
Here $P$, $A$ and $T$ are scalar form factors, the explicit form of which can be found in Appendix A of Ref.~\cite{Husek:2014tna}.

To get the NLO one-photon-irreducible part of the correction $\delta$ we need to consider the interference term of LO matrix element (\ref{eq:MD}) and the 1$\gamma$IR contribution (\ref{eq:M1gI}) and sum it over the photon polarizations with the result
\begin{equation}
\begin{split}
&\overline{\mathcal{M}_{1\gamma\text{IR}}^\text{LO}(x,y)}
\equiv\sum_\text{polar.}{\big[\mathcal{M}^{\text{LO}}(p,q,k)\big]^*\mathcal{M}_{1\gamma\text{IR}}}(p,q,k)\\
&=-\frac{ie^8M^3}{8}\frac{(1-x)^2}{x}\mathcal{F}^*(x)\mathcal{F}_{\pi^0\gamma\gamma}^\text{LO}
\Big\{4\nu T(x,y)\\
&+\Big[A(x,y)M[x(1-y)^2-\nu^2]+(y\to-y)\Big]\Big\}\,.
\end{split}
\end{equation}
Putting the above formula into (\ref{eq:dGammaxy}) and (\ref{eq:dxy}) and normalizing to the LO two-fold differential decay width (\ref{eq:dLOxy}) we get finally
\begin{equation}
\begin{split}
&\delta^{1\gamma\text{IR}}(x,y)
=2\operatorname{Re}\Big\{\overline{\mathcal{M}_{1\gamma\text{IR}}^\text{LO}(x,y)}\Big\}\frac{M(1-x)}{(8\pi)^3}
\bigg/\frac{{\mathrm d}^2\Gamma^\text{LO}}{{\mathrm d} x{\mathrm d} y}\\
&=2\operatorname{Re}\left\{-\frac{\alpha}{\pi}\frac{\mathcal{F}^\text{LO}(0)}{\mathcal{F}(x)}\frac{i\pi^2M}{\left[1+y^2+\frac{\nu^2}{x}\right]}
\Big\{4\nu T(x,y)\right.\\
&+\Big[A(x,y)M[x(1-y)^2-\nu^2]+(y\to-y)\Big]\Big\}\bigg\}\,.
\end{split}
\label{eq:d1gIR}
\end{equation}
For our purpose we can safely set $\mathcal{F}(x)\simeq\mathcal{F}^\text{LO}(0)$ in the previous formula, considering only the leading order of the chiral expansion; see also (\ref{eq:fx}) assuming the slope $a$ is small.
It should be mentioned, though, that such an approximation is only reasonable for the Dalitz decay of a neutral pion.
For the decays of an eta meson, one should be more cautious and use a better treatment of the full form factor.

Similarly, the dependence on the parameter $\chi^\text{(r)}$ cannot be neglected when $\nu$ becomes significant.
Indeed, considering the full expression (A.5) from~\cite{Husek:2014tna} for the form factor $T(x,y)$, one gets for the $\chi$-dependent contribution to $\delta^{1\gamma\text{IR}}(x,y)$ from (\ref{eq:d1gIR})
\begin{equation}
\delta_{\chi^\text{(r)}}^{1\gamma\text{IR}}(x,y)
=-\frac{\alpha}{\pi}\frac{\mathcal{F}^\text{LO}(0)}{\mathcal{F}(x)}
\frac{4\nu^2\chi^\text{(r)}(\mu)}{(1-x)(1-y^2)}\frac{1}{\left[1+y^2+\frac{\nu^2}{x}\right]}\,.
\label{eq:d1gIRchi}
\end{equation}
Thus, e.g.\ for the decay $\eta\to\mu^+\mu^-\gamma$ the one-photon-irreducible contribution may be considerably model-dependent.
This is, however, not the case for the process $\pi^0\to e^+e^-\gamma$ where the contribution given in (\ref{eq:d1gIRchi}) is suppressed in comparison to the other terms in (\ref{eq:d1gIR}).


\section{Bremsstrahlung}
\label{sec:BS}

\begin{figure}[!ht]
\centering
\setlength{\unitlength}{0.55pt}
\begin{picture}(0,180) (280,-200)
    \SetScale{0.55}
    \SetWidth{1.0}
    \SetColor{Black}
    \Line[dash,dashsize=10](70,-103)(130,-103)
    \Photon(206,-157)(134,-103){4}{5.5}
    \Line[arrow,arrowpos=0.5,arrowlength=5,arrowwidth=2,arrowinset=0.2](242,-103)(188,-67)
    \Line[arrow,arrowpos=0.3,arrowlength=5,arrowwidth=2,arrowinset=0.2](188,-67)(242,-31)
    \Photon(134,-103)(188,-67){4}{4.5}
    \Photon(221,-45)(245,-60){2}{4}
    \GOval(140,-103)(10,10)(0){0.882}
    \Line[dash,dashsize=10](314,-103)(374,-103)
    \Photon(450,-157)(378,-103){4}{5.5}
    \Line[arrow,arrowpos=0.7,arrowlength=5,arrowwidth=2,arrowinset=0.2](486,-103)(432,-67)
    \Line[arrow,arrowpos=0.5,arrowlength=5,arrowwidth=2,arrowinset=0.2](432,-67)(486,-31)
    \Photon(378,-103)(432,-67){4}{4.5}
    \Photon(465,-89)(489,-74){2}{4}
    \GOval(384,-103)(10,10)(0){0.882}
    \Text(240,-205)[lb]{\Large{\Black{$\mathrm{ }$}}}
    \Text(240,-195)[lb]{\Large{\Black{$\mathrm{+~cross}$}}}
\end{picture}
\label{fig:Bb}
\caption{
\label{fig:BS}
Bremsstrahlung corrections for $\pi^0\to e^+e^-\gamma$ process.
Needless to say, ``cross'' stands for the diagrams with outgoing photons interchanged.
}
\end{figure}
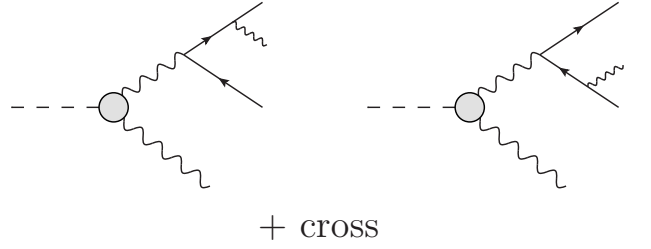

In this section we recapitulate the approach used in Ref.~\cite{Mikaelian:1972yg} for the bremsstrahlung correction calculation.
We think it is useful and convenient to rewrite the whole story in a more detailed way so it is transparent and easily understood.
As usual, one can then build on that when a few more pieces come into play.
In the Appendixes we then provide the results themselves.
Note also that especially in this section we restrict ourselves to the original notation used in the work~\cite{Mikaelian:1972yg}.

The diagrams which contribute to the Dalitz decay bremsstrahlung and are thus important to cancel the IR divergences stemming from the virtual corrections discussed in Sec.~\ref{sec:virt} are shown in Fig.~\ref{fig:BS}.
The corresponding invariant matrix element (including cross terms) can be written in the form
\begin{equation}
i\mathcal{M}_\text{BS}=\bar u(p)\big[I^{\rho\sigma}(k,l)+I^{\sigma\rho}(l,k)\big]v(q)\epsilon_\rho^*(k)\epsilon_\sigma^*(l)\,,
\label{eq:M}
\end{equation}
where\footnote{We use the shorthand notation for the product of the Levi-Civita tensor and four-momenta in which $\varepsilon^{(k)\dots}=\varepsilon^{\mu\dots} k_\mu$\,.}
\begin{equation}
\begin{split}
&I^{\alpha\beta}(k,l)
=-i^5e^4\mathcal{F}\bigg(\frac{(l+p+q)^2}{M^2}\bigg)\frac{\varepsilon^{(l+p+q)(k)\mu\alpha}}{(l+p+q)^2}\\
&\times\bigg[\gamma^\beta\frac{(\slashed l+\slashed p+m)}{2l\cdot p+i\epsilon}\gamma^\mu-\gamma^\mu\frac{(\slashed l+\slashed q-m)}{2l\cdot q+i\epsilon}\gamma^\beta\bigg]\,.
\end{split}
\end{equation}
The form factor $\mathcal{F}({(l+p+q)^2}/{M^2})$ can be expanded (assuming $a$ is small) in the following way
\begin{equation}
\mathcal{F}\bigg(\frac{(l+p+q)^2}{M^2}\bigg)
\simeq\mathcal{F}(x)\bigg[1+a\,\frac{2l\cdot(p+q)}{M^2}\bigg]\,.
\label{eq:BSFFexpand}
\end{equation}
Thus for the process $\pi^0\to e^+e^-\gamma$ it can be approximated by $\mathcal{F}(x)$, taking into account only the leading order in the chiral expansion.
Let us also introduce Tr for the rescaled matrix element squared and summed over all spins and polarizations of final states by the relation
\begin{equation}
\overline{|\mathcal{M}_\text{BS}|^2}
\equiv\sum_\text{sp., pol.}|\mathcal{M}_\text{BS}|^2
\equiv\frac{e^8}{4}|\mathcal{F}(x)|^2\,\text{Tr}\,.
\label{eq:Tr}
\end{equation}

Inasmuch as an additional photon comes into play it is convenient to introduce a new kinematic variable which describes the normalized invariant mass squared of the two photons
\begin{equation}
x_\gamma=\frac{(k+l)^2}{M^2}\,.
\end{equation}
It has the similar meaning as $x$ in the case of the electron-positron pair. The limits on $x_\gamma$ are
\begin{equation}
\frac{\lambda^2}{M^2}\le x_\gamma\le x_\gamma^\text{max}\equiv 1+x-\sqrt{4x+\frac{y^2}{\beta^2}(1-x)^2}\,.
\end{equation}

The contribution of the bremsstrahlung to the next-to-leading order can be described (according to (\ref{eq:dxy})) by the correction
\begin{equation}
\delta^\text{BS}(x,y)
=\frac{{\mathrm d}^2\Gamma^\text{BS}}{{\mathrm d} x{\mathrm d} y}\bigg/\frac{{\mathrm d}^2\Gamma^\text{LO}}{{\mathrm d} x{\mathrm d} y}\,,
\label{eq:deltaBS}
\end{equation}
in which in agreement with \cite{Mikaelian:1972yg} we can write
\begin{equation}
\begin{split}
\frac{{\mathrm d}^2\Gamma^\text{BS}}{{\mathrm d} x{\mathrm d} y}
&=\frac{(1-x)}{4M(2\pi)^8}\frac{\pi^3M^4}{16}\int J\Big[\overline{|\mathcal{M}_\text{BS}|^2}\Big]{\mathrm d}x_\gamma\\
&=\frac{|f(x)|^2}{64}\left(\frac{\alpha}{\pi}\right)^2\Gamma_{\pi^0\to\gamma\gamma}^\text{LO}(1-x)\int J[\text{Tr}]\,{\mathrm d}x_\gamma\,.
\end{split}
\label{eq:dBS}
\end{equation}
The above used operator $J$ is defined for an arbitrary invariant $f(k,l)$ of the momenta $k$ and $l$ as follows
\begin{equation}
J[f(k,l)]
=\frac 1{2\pi}\int\frac{{\mathrm d}^3k}{k_0}\frac{{\mathrm d}^3l}{l_0}f(k,l)\,\delta^{(4)}(P-p-q-k-l)\,.
\end{equation}
Finally, putting the LO differential decay width expression (\ref{eq:dLOxy}) and the previous result (\ref{eq:dBS}) into (\ref{eq:deltaBS}) we get
\begin{equation}
\delta^\text{BS}(x,y)
=\frac{1}{64}\left(\frac{\alpha}{\pi}\right)\frac{4x}{(1-x)^2}\frac{\int J[\text{Tr}]\,{\mathrm d}x_\gamma}{\left[1+y^2+\frac{\nu^2}{x}\right]}\,.
\label{eq:dBSxy}
\end{equation}
In the remaining part of this section we discuss the way the integral $\int J[\text{Tr}]\,{\mathrm d}x_\gamma$ is treated. Most of the explicit formulas are then moved to the Appendixes.

Being on shell ($k^2=0=l^2$) and in the diphoton center-of-mass system where $\vv{P}-\vv{p}-\vv{q}=0(=\vv{k}+\vv{l}\equiv\vv{r})$, we find
\begin{equation}
\begin{split}
J[f(k,l)]
\stackrel{(\vv{r}=0)}{=}\frac 1{4\pi}\int{\mathrm d}\Omega\,f(k,\tilde{k})
=\frac 1{4\pi}\int{\mathrm d}\Omega\,f(\tilde{l},l)\,.
\end{split}
\label{eq:Jint}
\end{equation}
Here, we have used $\tilde{l}$ to mark the four-momentum $l$ with the opposite momentum direction, i.e.\ whenever $l=(l_0,\vv l)$, then $\tilde l=(l_0,-\vv l)$\,.
We can come back to the invariant form in a known way through
\begin{equation}
2l_0=k_0+l_0
\stackrel{(\vv{r}=0)}{=}\sqrt{(k+l)^2}=M\sqrt{x_\gamma}
\end{equation}
or, for example, due to
\begin{equation}
p_0=\frac{(k_0+l_0)p_0}{(k_0+l_0)}
\stackrel{(\vv{r}=0)}{=}\frac{(k+l)\cdot p}{M\sqrt{x_\gamma}}\,.
\end{equation}

If we follow the notation of Ref.~\cite{Mikaelian:1972yg}, we define the propagator denominators in the following way (suppressing +$i\epsilon$ part for now)
\begin{equation}
\begin{gathered}
A=l\cdot q\,,\quad
B=l\cdot p\,,\quad
C=k\cdot q\,,\quad
D=k\cdot p\,,\quad\\
E=(p+q+l)^2\,,\quad
F=(p+q+k)^2\,.
\end{gathered}
\label{eq:ABCDEF}
\end{equation}
Not only is the whole amplitude invariant under the interchange of the two photons (and thus of $k$ and $l$ in (\ref{eq:M})), but also the operator $J$ possesses the same symmetry which can be written for an arbitrary function of the propagator denominators (\ref{eq:ABCDEF}) as
\begin{equation}
J[f(A,B,C,D,E,F)]=J[f(C,D,A,B,F,E)]\,.
\label{eq:sym1}
\end{equation}
The interchange of $p$ and $q$ (which is also a relevant symmetry in our case) must be compensated on the level of the operator $J$ by changing the $y$ sign, thus
\begin{equation}
J[f(A,B,C,D,E,F)]
=J[f(B,A,D,C,E,F)]\Big|_{y\to-y}\,.
\label{eq:sym2}
\end{equation}

There are also some useful identities which follow from the definitions (\ref{eq:ABCDEF}) such as
\begin{align}
\hspace{-2mm}E-2A-2B
=F-2C-2D
&=M^2x\,,
\label{eq:EAB}\\
\hspace{-2mm}
F+2A+2B
=E+2C+2D
&=M^2(1-x_\gamma)
\label{eq:ECD}
\end{align}
and
\begin{align}
A+C&=\frac{M^2}4\left[(1-x)(1+y)-x_\gamma\right]\,,\label{eq:AC}\\
B+D&=\frac{M^2}4\left[(1-x)(1-y)-x_\gamma\right]\,,\\
&=(A+C)\Big|_{y\to-y}\notag\\
E+F&=M^2(1+x-x_\gamma)\,.
\end{align}
It is convenient to know the above relations for two reasons.
First, we see that we can simply trade one of the above defined variables for the others and thus only two more independent variables in addition to $x$, $y$ and $x_\gamma$ (e.g.\ $A$ and $B$) are necessary to describe the kinematics of our decay.
On the other hand, we realize that some special combinations of the variables $A,\,\dots, F$ are invariant with respect to the acting of the operator $J$ (i.e.\ they depend only on $x$, $y$ and $x_\gamma$).
We can also combine the previous formulas to get some other $J$-invariant combinations. If we consider, for example, that
\begin{equation}
A
=(A+C)-\frac 12 [(E+2C+2D)-E-2D]\,,
\end{equation}
we find
\begin{equation}
\frac E2-A+D=\frac{M^2}4[1+x-x_\gamma-y(1-x)]\,.
\label{eq:ADE}
\end{equation}

Such expressions are useful when we want to reduce the complicated $J$ terms, arising naturally during the calculation of the invariant matrix element squared, to the basic ones which are simple to handle.
First, we use the above stated relations to simplify the numerators (e.g.\ we get rid of $A$ in a term like $A$/($DE$) using the relation (\ref{eq:ADE})).%
\footnote{The combination (\ref{eq:ADE}) is of course in some minimalistic sense redundant for the considered procedure, since we can always make two-step substitution instead. In such a case, we would trade $A$ for $C$ using (\ref{eq:AC}) and then $C$ for $E$ and $D$ using (\ref{eq:ECD}).}
Then also the denominators are treated.
For example, consider the term $J[1/(ACEF)]$.
Then
\begin{equation}
\begin{split}
&\frac 1{ACEF}
=\frac 1{(A+C)}\frac 1{(E+F)}\frac{(A+C)(E+F)}{ACEF}\\
&=\frac 1{(A+C)(E+F)}\left(\frac{1}{AE}+\frac{1}{AF}+\frac{1}{CE}+\frac{1}{CF}\right)\,.
\end{split}
\end{equation}
After applying the operator $J$ and using the symmetry (\ref{eq:sym1}), we find
\begin{equation}
\begin{split}
J\bigg[\frac 1{ACEF}\bigg]&=\frac 2{(A+C)(E+F)}\\
&\times\left(J\bigg[\frac 1{AE}\bigg]+J\bigg[\frac 1{CE}\bigg]\right)\,.
\end{split}
\end{equation}
All necessary reductions of this type are summarized in Appendix \ref{app:J}, except for such terms which one can get using the discussed symmetries (\ref{eq:sym1}) and (\ref{eq:sym2}).
The computational methods used to calculate the basic terms are introduced in Appendix \ref{app:meth}.
For the list of the results for these integrals see Appendix \ref{app:Jbasic}.
Here, in comparison to Ref.~\cite{Mikaelian:1972yg}, we include also the new term $J[1/(A^2E^2)]$ which appears due to the fact that $\mathcal{O}(\nu^4)$ terms were not neglected in our approach.
The completely reduced rescaled matrix element squared Tr, which represents in terms of $J[\text{Tr}]$ an important ingredient for the bremsstrahlung correction $\delta^\text{BS}(x,y)$ (cf.~(\ref{eq:dBSxy})), is presented in Appendix \ref{app:M}.
We believe we provide here the results in a more refined way in comparison with Ref.~\cite{Mikaelian:1972yg}.

The last step is the integration over~$x_\gamma$.
There are basic integrals which behave like 1/$x_\gamma$ and are divergent when this integration is performed if no $x_\gamma$ appears in the numerator to compensate it.
The essential divergent integrals are $J[1/A^2]$ and $J[1/(AB)]$.
The divergent part of integrals like $J[1/(A^2E)]$ and $J[1/(A^2E^2)]$ can then be written in terms of these essential ones.
For example, using (\ref{eq:EAB}) we get
\begin{equation}
\frac1{A^2E}
=\frac1{M^2x}\left(\frac1{A^2}-\frac2{AE}-\frac{2B}{A^2E}\right).
\end{equation}
Needless to say, there are also $A\to B$ counterparts of the mentioned integrals.
This unwelcome behavior can be extracted from the Tr expression to get the convergent part Tr$_\text{C}$, which can be treated numerically, and the divergent part Tr$_\text{D}$, which should be treated analytically.
In the former case we can set $\lambda\to 0$ and the lower bound on $x_\gamma$ is then zero.
In the latter case the cutoff $\lambda$ has to be preserved.

Finally, as expected, the sum of the divergent part of the bremsstrahlung correction $\delta^\text{BS}_\text{D}(x,y)$, the explicit form of which can be found in (\ref{eq:deltaBSD}), and the divergent part of virtual correction $\delta^\text{virt}(x,y)$, represented in the following formula by the electromagnetic form factor $F_1(x)$, in particular
\begin{equation}
\delta^\text{BS}_\text{D}(x,y)+2\operatorname{Re}\big\{{F_1}(x)\big\}\,,
\end{equation}
is IR finite. In other words, terms proportional to $\log m/\lambda$ cancel each other in the final formula of the correction $\delta(x,y)$.

In the end of this section, let us go back to Eq. (\ref{eq:BSFFexpand}).
In cases when the slope $a$ is no longer negligible in comparison to 1, one should consider the entire right-hand side of (\ref{eq:BSFFexpand}) instead of only $\mathcal{F}(x)$ alone.
It is then necessary to go beyond the approach used in Ref.~\cite{Mikaelian:1972yg}.
If we square the bremsstrahlung matrix element (\ref{eq:M}), we get for the simple case with $a=0$ which we have treated so far
\begin{equation}
{|\mathcal{M}_\text{BS}^{a=0}|^2}
={|I_{k,l}+I_{l,k}|^2}
=|I_{k,l}|^2+|I_{l,k}|^2+2I_{k,l}^*I_{l,k}^{}\,.
\label{eq:MBSI}
\end{equation}
Here we have denoted
\begin{equation}
I_{k,l}
\equiv\bar u(p)I_{a=0}^{\rho\sigma}(k,l)v(q)\epsilon_\rho^*(k)\epsilon_\sigma^*(l)
\end{equation}
and likewise for $I_{l,k}$.
Using the building blocks of the ``no-slope'' matrix element modulus squared (\ref{eq:MBSI}) and considering the expansion (\ref{eq:BSFFexpand}) we find the correction for the bremsstrahlung expression
\begin{equation}
\begin{split}
\overline{|\mathcal{M}_\text{BS}^{a\ne 0}|^2}
&=[1+a(1-x-x_\gamma)]\overline{|\mathcal{M}_\text{BS}^{a=0}|^2}\\
&+a\frac{(E-F)}{M^2}\left(\overline{|I_{k,l}|^2}-\overline{|I_{l,k}|^2}\right)\,.
\end{split}
\end{equation}
If we apply the operator $J$ and take into account the symmetry (\ref{eq:sym1}), the previous formula can be boiled down to
\begin{equation}
\begin{split}
&J\Big[\overline{|\mathcal{M}_\text{BS}^{a\ne 0}|^2}-\overline{|\mathcal{M}_\text{BS}^{a=0}|^2}\Big]\\
&=2a\bigg\{(1-x-x_\gamma)\,J\Big[\overline{I_{k,l}^*I_{l,k}^{}}\Big]+4\,J\bigg[\frac{(A+B)}{M^2}\overline{|I_{k,l}|^2}\bigg]\bigg\}\,.\\
\end{split}
\label{eq:MBSa}
\end{equation}
This expression can be calculated along the same lines as $J[\text{Tr}]$.
One then gets a similar expression to Tr$_\text C$ in (\ref{eq:TrC}) including some new integrals.
These need to be calculated in addition to the known basic terms.
Note that there is no divergent part in (\ref{eq:MBSa}) which needs to be treated separately.

The above correction does not need to be considered in the decay $\pi^0\to e^+e^-\gamma$ so we do not present the related results in this paper.
On the other hand, it becomes important when treating the eta meson decays.


\section{Results}
\label{sec:res}

For the reader's convenience, we put here together the individual pieces (\ref{eq:dvirt}), (\ref{eq:d1gIR}) and (\ref{eq:dBSxy}) and write the overall NLO correction 
\begin{equation}
\begin{split}
&\delta(x,y)
=\delta^\text{virt}(x,y)+\delta^{1\gamma\text{IR}}(x,y)+\delta^\text{BS}(x,y)\\
&=2\operatorname{Re}\left\{-\Pi(x)+F_1(x)+\frac{2F_2(x)}{1+y^2+\frac{\nu^2}{x}}\right.\\
&-\left(\frac{\alpha}{\pi}\right)\frac{M}{\left[1+y^2+\frac{\nu^2}{x}\right]}(i\pi^2)
\Big\{4\nu T(x,y)\\
&+\Big[A(x,y)M[x(1-y)^2-\nu^2]+(y\to-y)\Big]\Big\}
\Bigg\}\\
&+\frac{1}{64}\left(\frac{\alpha}{\pi}\right)\frac{4x}{(1-x)^2}\frac{\int J[\text{Tr}_\text C]{\mathrm d}x_\gamma}{\left[1+y^2+\frac{\nu^2}{x}\right]}
+\delta^\text{BS}_\text{D}(x,y)\,.
\end{split}
\label{eq:dtotalxy}
\end{equation}
Here, the convergent part of the rescaled bremsstrahlung invariant matrix element squared (to be integrated over $x_\gamma$ numerically) $\text{Tr}_\text{C}$ is given by (\ref{eq:TrC}) and the analytically integrated divergent part of the bremsstrahlung correction $\delta^\text{BS}_\text{D}(x,y)$ is shown in (\ref{eq:deltaBSD}).
Let us recall that the explicit formulas for the scalar form factors $A$ and $T$ can be found in Appendix A of Ref.~\cite{Husek:2014tna}.

\begin{figure}[t!]
\includegraphics[width=\columnwidth]{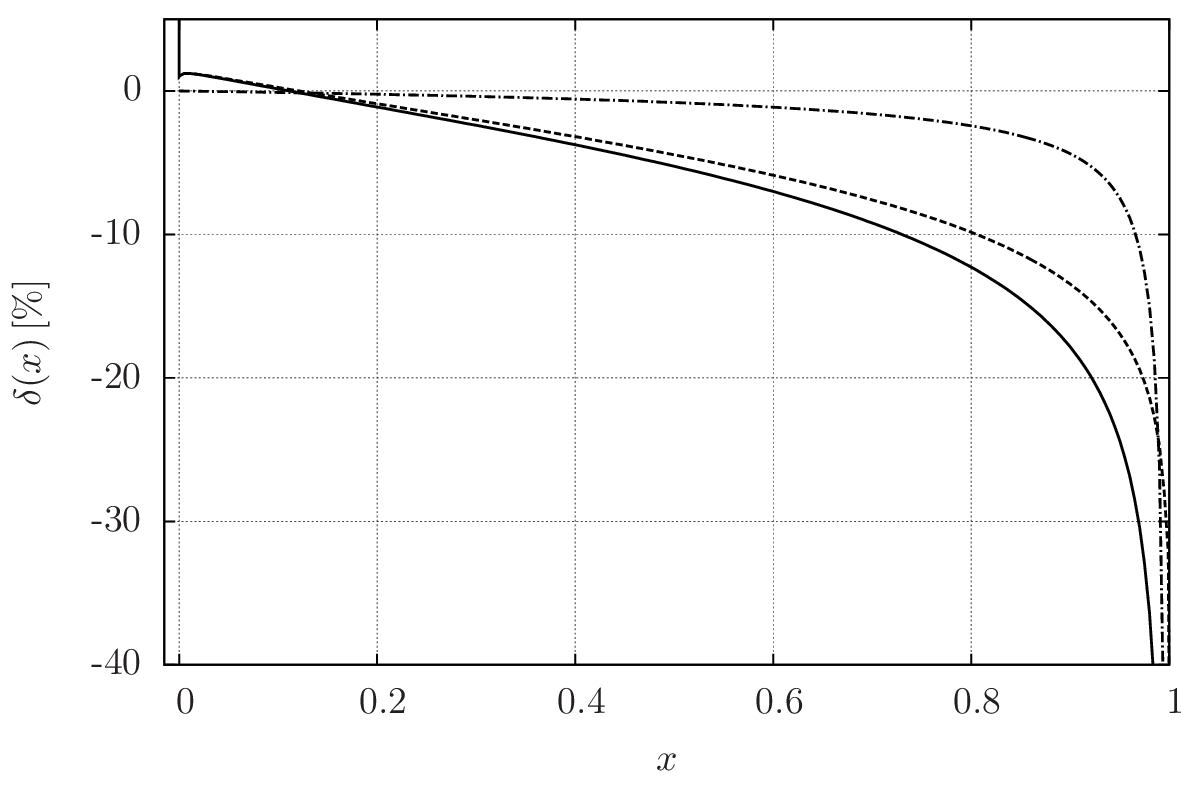}
\caption{\label{fig:dx2D}
The overall NLO correction $\delta(x)$ for the decay $\pi^0\to e^+e^-\gamma$ calculated according to the formula (\ref{eq:dtotalxy}) (solid line) in comparison to its constituents.
The sum $\delta^\text{virt}(x)+\delta^\text{BS}(x)$ is depicted as a dashed line and the one-photon-irreducible contribution $\delta^{1\gamma\text{IR}}$ is shown as a dash-dot line.
The divergent behavior of $\delta(x)$ near $x=\nu^2\simeq0$ has the origin in the electromagnetic form factor $F_1(x)$ and is connected to the Coulomb self-interaction of the dilepton at the threshold.
}
\end{figure}

Taking the result (\ref{eq:dtotalxy}) and using the formula (\ref{eq:dx}), we get the overall correction to the one-fold differential leading-order decay width, which is shown in Fig.~\ref{fig:dx2D}.
For comparison, also the sum $\delta^\text{virt}(x)+\delta^\text{BS}(x)$, which would have corresponded to the correction presented in the original paper~\cite{Mikaelian:1972yg} if the $\mathcal{O}(\nu^4)$ terms and the muon loop had not been omitted, and one-photon-irreducible contribution $\delta^{1\gamma\text{IR}}$ are shown.
We see that in the case of the decay $\pi^0\to e^+e^-\gamma$ the 1$\gamma$IR correction is negative for the whole range of values of $x$ and enhances thus the effect of the sum $\delta^\text{virt}(x)+\delta^\text{BS}(x)$ which is also negative in a wide range of $x$.

Taking into account all the discussed contributions, a similar table of values of correction $\delta(x,y)$ as it was provided in the original work~\cite{Mikaelian:1972yg}, can be produced at the very same points according to (\ref{eq:dtotalxy}); see Table~\ref{tab:MSpiee}.
Considering the contributions introduced in this work but left out in Ref.~\cite{Mikaelian:1972yg}, the 1$\gamma$IR correction is the most important one, especially for large $x$.
The correction of the old Mikaelian and Smith values is significant and greater than 10\,\% already for $x\simeq0.5$. 
This can be visible in Fig.~\ref{fig:dx2D} and also from the difference of the entry values between the Table I in Ref.~\cite{Mikaelian:1972yg} and Table~\ref{tab:MSpiee} in the present work, provided the remaining contributions are not significant.
Indeed, the muon loop vacuum polarization insertion contribution, which is independent on $y$, grows nearly linearly with $x$ from $\delta_{\mu\text{-loop}}^\text{virt}(0.01,y)=-0.0005$\,\% up to $\delta_{\mu\text{-loop}}^\text{virt}(0.99,y)=-0.0616$\,\% and is thus negligible.
A similar conclusion holds then also for the $\mathcal{O}(\nu^4)$ contribution, which is most significant for small $x$ with the value $\delta_{\nu^4}^\text{BS}(0.01,0)=0.0035$\,\%.

With our present knowledge, we are now in a position to calculate the correction to the integrated decay width.
In this case, the transition form factor $\mathcal{F}(x)$ cannot be scaled out anymore.
On the other hand, for relevant examples~\cite{Husek:2015wta} this model dependence is negligible for the decay $\pi^0\to e^+e^-\gamma$ and we get $\delta=8.30\times10^{-3}$.
This can be rewritten in a common way as
\begin{equation}
\frac{\Gamma_{\pi^0\to e^+e^-\gamma}^\text{NLO}}{\Gamma_{\pi^0\to\gamma\gamma}^\text{LO}}
=0.986\times10^{-4}\,.
\end{equation}
Without the inclusion of the 1$\gamma$IR contribution, the above number would become $1.03\times10^{-4}$.
The stated values are consistent with the previous results $1.05\times10^{-4}$ of Joseph~\cite{Joseph:1960zz} and $0.95\times10^{-4}$ of Mikaelin and Smith, who admitted that eventual numerical inaccuracy might be present in their result~\cite{Mikaelian:1972yg}.

\begingroup
\squeezetable
\begin{table*}[t]
\caption{
\label{tab:MSpiee}
The overall NLO correction $\delta(x,y)$ given in percent for a range of values of $x$ and $y$ (i.e.\ the Dalitz-plot corrections) for the process $\pi^0\to e^+e^-\gamma$.
}
\begin{ruledtabular}
\begin{tabular}{l d d d d d d d d d d d}
\, \backslashbox{x}{y} & 0.00 & 0.10 & 0.20 & 0.30 & 0.40 & 0.50 & 0.60 & 0.70 & 0.80 & 0.90 & 0.99 \\
\hline
0.01 & 2.761 & 2.714 & 2.599 & 2.449 & 2.273 & 2.061 & 1.786 & 1.402 & 0.803 & -0.357 & -5.657\rule{0pt}{3.5ex}\\
0.02 & 2.756 & 2.720 & 2.622 & 2.480 & 2.300 & 2.073 & 1.774 & 1.355 & 0.703 & -0.546 & -5.859 \\
0.03 & 2.669 & 2.639 & 2.552 & 2.419 & 2.242 & 2.012 & 1.704 & 1.267 & 0.586 & -0.716 & -6.125 \\
0.04 & 2.558 & 2.531 & 2.452 & 2.327 & 2.155 & 1.925 & 1.611 & 1.164 & 0.464 & -0.874 & -6.372 \\
0.05 & 2.437 & 2.412 & 2.340 & 2.221 & 2.053 & 1.824 & 1.509 & 1.054 & 0.341 & -1.025 & -6.601 \\
0.06 & 2.311 & 2.288 & 2.221 & 2.108 & 1.944 & 1.717 & 1.400 & 0.940 & 0.216 & -1.172 & -6.815 \\
0.07 & 2.184 & 2.163 & 2.099 & 1.990 & 1.830 & 1.605 & 1.288 & 0.824 & 0.092 & -1.315 & -7.017 \\
0.08 & 2.056 & 2.036 & 1.975 & 1.870 & 1.714 & 1.491 & 1.173 & 0.707 & -0.033 & -1.455 & -7.211 \\
0.09 & 1.928 & 1.909 & 1.851 & 1.749 & 1.596 & 1.374 & 1.057 & 0.588 & -0.157 & -1.593 & -7.397 \\
0.10 & 1.801 & 1.783 & 1.726 & 1.628 & 1.477 & 1.257 & 0.940 & 0.469 & -0.281 & -1.729 & -7.578\rule{0pt}{3.5ex}\\
0.15 & 1.170 & 1.154 & 1.105 & 1.016 & 0.874 & 0.661 & 0.345 & -0.131 & -0.900 & -2.394 & -8.424 \\
0.20 & 0.546 & 0.532 & 0.486 & 0.402 & 0.266 & 0.057 & -0.258 & -0.738 & -1.520 & -3.048 & -9.219 \\
0.25 & -0.079 & -0.092 & -0.135 & -0.217 & -0.350 & -0.556 & -0.871 & -1.355 & -2.148 & -3.704 & -9.995 \\
0.30 & -0.713 & -0.726 & -0.768 & -0.847 & -0.978 & -1.184 & -1.499 & -1.988 & -2.790 & -4.372 & -10.770 \\
0.35 & -1.366 & -1.378 & -1.419 & -1.497 & -1.627 & -1.833 & -2.149 & -2.641 & -3.454 & -5.058 & -11.558 \\
0.40 & -2.044 & -2.056 & -2.097 & -2.174 & -2.304 & -2.509 & -2.827 & -3.324 & -4.146 & -5.773 & -12.370 \\
0.45 & -2.759 & -2.771 & -2.811 & -2.887 & -3.017 & -3.222 & -3.543 & -4.044 & -4.875 & -6.525 & -13.218 \\
0.50 & -3.521 & -3.533 & -3.572 & -3.648 & -3.777 & -3.983 & -4.306 & -4.811 & -5.653 & -7.324 & -14.115 \\
0.55 & -4.344 & -4.356 & -4.395 & -4.470 & -4.599 & -4.806 & -5.130 & -5.640 & -6.492 & -8.186 & -15.076\rule{0pt}{3.5ex}\\
0.60 & -5.249 & -5.261 & -5.299 & -5.373 & -5.501 & -5.708 & -6.034 & -6.549 & -7.410 & -9.128 & -16.123 \\
0.65 & -6.262 & -6.273 & -6.310 & -6.383 & -6.510 & -6.717 & -7.044 & -7.563 & -8.435 & -10.177 & -17.284 \\
0.70 & -7.425 & -7.435 & -7.470 & -7.541 & -7.666 & -7.871 & -8.198 & -8.721 & -9.603 & -11.371 & -18.602 \\
0.75 & -8.802 & -8.811 & -8.844 & -8.910 & -9.031 & -9.232 & -9.558 & -10.084 & -10.976 & -12.772 & -20.143 \\
0.80 & -10.508 & -10.516 & -10.544 & -10.604 & -10.717 & -10.912 & -11.233 & -11.759 & -12.659 & -14.486 & -22.024 \\
0.85 & -12.779 & -12.784 & -12.804 & -12.851 & -12.949 & -13.129 & -13.438 & -13.958 & -14.864 & -16.724 & -24.468 \\
0.90 & -16.207 & -16.205 & -16.206 & -16.225 & -16.289 & -16.434 & -16.712 & -17.208 & -18.108 & -20.003 & -28.003 \\
0.95 & -23.167 & -23.144 & -23.084 & -23.011 & -22.960 & -22.982 & -23.140 & -23.532 & -24.360 & -26.256 & -34.451 \\
0.99 & -54.287 & -54.068 & -53.442 & -52.496 & -51.351 & -50.147 & -49.029 & -48.155 & -47.761 & -48.467 & -55.831\rule[-1.5ex]{0pt}{0pt}\\
\end{tabular}
\end{ruledtabular}
\end{table*}
\endgroup


\section{Summary}
\label{sec:sum}

In the preceding sections we have explored all the relevant NLO radiative corrections to the Dalitz decay of a neutral pion in the QED sector.
In the direct comparison to the earlier approach of Mikaelian and Smith~\cite{Mikaelian:1972yg}, we have included into our treatment the one-photon-irreducible contribution.
On the top of that, as announced above we have enriched the vacuum polarization insertion correction with the muon loop and have not thrown away the $\mathcal{O}(\nu^4)$ terms.
The latter is connected to the calculation of an additional nontrivial integral.
On the other hand, we were able to write the results in a more compact form even though more terms needed to be covered.
The computational methods as well as some intermediate results are also provided and thus it should be possible for an interested reader to trace back all the steps made.

From the newly included contributions only the 1$\gamma$IR correction is relevant for the decay $\pi^0\to e^+e^-\gamma$ and should be introduced in the future analyses.
Needless to say, the provided calculation is universal considering the masses of the particles involved.
It can thus be also shown via direct calculation that if we change the masses of the particles in such a way that they correspond to the process $\eta\to \mu^+\mu^-\gamma$, all the discussed corrections should be taken into account.
In other words, both the muon loop as well as the $\mathcal{O}(\nu^4)$ terms give then a non-negligible contribution to the overall $\delta(x,y)$.
That is why these corrections should not be overlooked.
If necessary, heavier charged fermions may be also introduced in the loops in the same way the muon loop was added.

We believe that this work is a good starting point for a treatment of some other processes such as the Dalitz decays of $\eta$.
We have also touched on some particular difficulties that appear and one needs to be careful about.
A more detailed review of this matter is beyond the scope of this work and will be discussed separately in the paper in preparation.

Let us also say that after the complete recalculation of the results given in Ref.~\cite{Mikaelian:1972yg} we have verified the formulas therein.
The numerical accuracy of the listed values is also sufficient.

The main message of the present work is the completion of the list of the NLO corrections and refining of the expressions.
All the formulas necessary for the calculation of the considered correction are listed in the present paper in a ready-to-use form.
For the eventual future practical use of an interested reader, we submit together with this text also (as ancillary files) a \verb!C++! code, which contains all the expressions in a well-arranged way.
As a demonstration, the resulting program calculates the correction $\delta(x,y)$.

\section*{Acknowledgment}
We would like to thank Evgueni Goudzovski for turning our attention to this problem.

This work was supported by Charles University in Prague (Grant No.~GAUK 700214), by Ministry of Education of the Czech Republic (Grant No.~LG 13031) and by the Czech Science Foundation (Grant No.~GA\v{C}R 15-18080S).


\onecolumngrid
\newpage
\appendix


\section{Bremsstrahlung matrix element squared}
\label{app:M}

For the rescaled bremsstrahlung invariant matrix element squared Tr (see (\ref{eq:Tr}) for the definition) we can write
\begin{equation}
J[\text{Tr}]=J[\text{Tr}_\text C]+J[\text{Tr}_\text D]\,.
\end{equation}
The explicit forms of the (from the point of view of the integration over $x_\gamma$) convergent part Tr$_\text C$ and the divergent part Tr$_\text D$ of Tr are shown below%
\footnote{Note that $\{y\to-y\}$ in Tr$_\text C$ holds for the entire expression (including terms independent of $y$).}.
The terms are already reduced (see Appendix \ref{app:J} for the reduction procedure) to the basic integrals (see Appendix \ref{app:Jbasic} for the explicit expressions) and the symmetries of the operator $J$ (\ref{eq:sym1}) and (\ref{eq:sym2}) were used. This means the relation Tr\,=\,Tr$_\text C$\,+\,Tr$_\text D$ holds only effectively (with operator $J$ applied).

\begin{equation}
\begin{split}
\text{Tr}_\text C
&=16-4M^2\left[(1-x)(3-y)-3x_\gamma+3\nu^2+\frac{2\nu^2(x+x_\gamma)}{(1-x)(1+y)-x_\gamma}\right]\frac 1{A}\\
&-\frac{\nu^2M^4}x\frac {x_\gamma^2}{A^2}+M^4\left(2+\frac{\nu^2}{x}\right)\frac {x_\gamma^2}{AB}+16\frac BA-4\nu^2M^2\frac B{A^2}-16M^4\frac 1{E^2}-\frac{32M^2}{(1+x-x_\gamma)}\frac 1{E}\\
&+\frac{M^4}{2}\left[(1-x-x_\gamma)^2-4xx_\gamma-(1-x)^2y^2-2\nu^2(x-3x_\gamma)+\frac{8\nu^2(x-x_\gamma)^2}{(1+x-x_\gamma)^2-y^2(1-x)^2}\right]\frac 1{AD}\\
&+\frac{8\nu^2M^4(x-x_\gamma)^2}{(1+x-x_\gamma)[1+x-x_\gamma-y(1-x)]}\left(\frac 1{AE}-\frac 1{DE}\right)\\
&-4M^4\left\{2(1-x)^2(1+y^2)+2x_\gamma^2+[(1-x)^2(1-y^2)+x_\gamma^2]\frac{\nu^2}x-(1-2 x)\frac{\nu^4}{x^2}\right\}\frac 1{AE}\\
&-4M^4[(1-x-x_\gamma)^2+(1-x)^2y^2+2\nu^2(1+x_\gamma)]\frac 1{CE}-8M^6\left(x+\nu^2-\frac{\nu^4}{2x}\right)\frac 1{AE^2}\\
&-\nu^2M^6\{1+[(1-x)y-x+x_\gamma]^2-2\nu^2\}\left[\frac 1{A^2E}-\frac 1{M^2x}\frac 1{A^2}\right]-\nu^4M^8\left[\frac 1{A^2E^2}-\frac 1{M^4x^2}\frac 1{A^2}\right]\\
&+\frac{4M^4}{(1+x-x_\gamma)}\left(\frac 1{AE}+\frac 1{CE}\right)\bigg\{
(1+y^2)[2-x+x^3-2x_\gamma(1-x)^2]+2y[1-x-xy-x_\gamma(1-x^2)]\\
&-x_\gamma[6x+xx_\gamma+2(1-x_\gamma)^2]+2\nu^2[1+2x+x_\gamma(1+x-x_\gamma)]-\frac{4xx_\gamma(x^2+x_\gamma^2)-2\nu^2(x+x_\gamma)^2}{[(1-x)(1+y)-x_\gamma]}\bigg\}\\
&+\{y\to-y\}
\end{split}
\label{eq:TrC}
\end{equation}

\begin{equation}
\begin{split}
\text{Tr}_\text D
&=M^4(1-x)^2\left(1+y^2+\frac{\nu^2}{x}\right)\left[4\left(1-\frac{\nu^2}{2x}\right)\frac 1{AB}-\frac{\nu^2}{x}\left(\frac 1{A^2}+\frac 1{B^2}\right)\right]
\end{split}
\end{equation}

The integration over $x_\gamma$ of $J[\text{Tr}_\text D]$ has to be done analytically. After substituting the appropriate expressions from (\ref{eq:JA2div}) and (\ref{eq:JABdiv}) and putting the result into (\ref{eq:dBSxy}), we find for the contribution of the divergent part to the bremsstrahlung correction
\begin{equation}
\delta^\text{BS}_\text{D}(x,y)
=(-2)\left(\frac{\alpha}{\pi}\right)
\left\{
\left(1+\frac{1+\beta^2}{2\beta}\log\gamma\right)\left[\log\frac{m}{\lambda}+\log\frac{2x_\gamma^\text{max}}{(1-x)}\right]
-\frac 12\log(1-y^2)-\frac{1+\beta^2}{4\beta}K(x,y)
\right\}\,,
\label{eq:deltaBSD}
\end{equation}
where $K(x,y)$ is given by (\ref{eq:Kxy}).
It is apparent that the IR divergent part indeed cancels with its counterpart in the virtual correction $\delta^\text{virt}(x,y)$\,.


\newpage

\section{Reduction of \texorpdfstring{$J$}{J} terms}
\label{app:J}

In this Appendix we summarize all the necessary reductions of the $J$ terms to the basic integrals, the results of which can be found in Appendix \ref{app:Jbasic}. The following formulas are used to get the matrix element squared in the form shown in Appendix \ref{app:M}. 

\begin{align}
J\bigg[\frac 1{EF}\bigg]
&=\frac 2{(E+F)}\,J\bigg[\frac 1{E}\bigg]
=\frac 2 {M^2(1+x-x_\gamma)}\,J\bigg[\frac 1{E}\bigg]\\
J\bigg[\frac A{EF}\bigg]
&=\frac {(A+C)}{(E+F)}\,J\bigg[\frac 1{E}\bigg]
=\frac{\left[(1-x)(1+y)-x_\gamma\right]}{4(1+x-x_\gamma)}\,J\bigg[\frac 1{E}\bigg]\\
J\bigg[\frac 1{AC}\bigg]
&=\frac 2{(A+C)}\,J\bigg[\frac 1A\bigg]
=\frac8{M^2\left[(1-x)(1+y)-x_\gamma\right]}\,J\bigg[\frac 1A\bigg]\\
J\bigg[\frac 1{BD}\bigg]
&=\frac 2{(B+D)}\,J\bigg[\frac 1B\bigg]
=\frac8{M^2\left[(1-x)(1-y)-x_\gamma\right]}\,J\bigg[\frac 1A\bigg]_{y\to-y}
=J\bigg[\frac 1{AC}\bigg]_{y\to-y}\\
\notag\\
J\bigg[\frac 1{AEF}\bigg]
&=\frac 1{(E+F)}\left(J\bigg[\frac 1{AE}\bigg]+J\bigg[\frac 1{CE}\bigg]\right)
=\frac 1{M^2(1+x-x_\gamma)} \left(J\bigg[\frac 1{AE}\bigg]+J\bigg[\frac 1{CE}\bigg]\right)\\
\begin{split}
J\bigg[\frac 1{ABE}\bigg]
&=\frac 1{(E-2A-2B)}\left(J\bigg[\frac 1{AB}\bigg]-2\,J\bigg[\frac 1{AE}\bigg]-2\,J\bigg[\frac 1{BE}\bigg]\right)\\
&=\frac 2{M^2x}\left(\frac 12\,J\bigg[\frac 1{AB}\bigg]-\,J\bigg[\frac 1{AE}\bigg]-\,J\bigg[\frac 1{AE}\bigg]_{y\to-y}\right)
\end{split}\\
\begin{split}
J\bigg[\frac 1{ABE^2}\bigg]
&=\frac 1{(E-2A-2B)}\left(J\bigg[\frac 1{ABE}\bigg]-2\,J\bigg[\frac 1{AE^2}\bigg]-2\,J\bigg[\frac 1{BE^2}\bigg]\right)\\
&=\left\{\frac 2{M^4x^2}\left(\frac 14\,J\bigg[\frac 1{AB}\bigg]-J\bigg[\frac 1{AE}\bigg]\right)-\frac 2{M^2x}\,J\bigg[\frac 1{AE^2}\bigg]\right\}+\{y\to-y\}
\end{split}\\
\notag\\
J\bigg[\frac 1{ACE}\bigg]
&=\frac 1{(A+C)}\left(J\bigg[\frac 1{AE}\bigg]+J\bigg[\frac 1{CE}\bigg]\right)
=\frac4{M^2\left[(1-x)(1+y)-x_\gamma\right]} \left(J\bigg[\frac 1{AE}\bigg]+J\bigg[\frac 1{CE}\bigg]\right)\\
\begin{split}
J\bigg[\frac 1{ADE}\bigg]
&=\frac 1{(\frac E2-A+D)}\left(\frac 12\,J\bigg[\frac 1{AD}\bigg]+J\bigg[\frac 1{AE}\bigg]-J\bigg[\frac 1{DE}\bigg]\right)\\
&=\frac 4{M^2[1+x-x_\gamma-y(1-x)]}\left(\frac 12\,J\bigg[\frac 1{AD}\bigg]+J\bigg[\frac 1{AE}\bigg]-J\bigg[\frac 1{CE}\bigg]_{y\to-y}\right)
\end{split}\\
J\bigg[\frac 1{ACEF}\bigg]
&=\frac 2{(A+C)}\,J\bigg[\frac 1{AEF}\bigg]
=\frac 8{M^4(1+x-x_\gamma)\left[(1-x)(1+y)-x_\gamma\right]} \left(J\bigg[\frac 1{AE}\bigg]+J\bigg[\frac 1{CE}\bigg]\right)\\
\begin{split}
J\bigg[\frac 1{ADEF}\bigg]
&=\frac 1{(E+F)}\left(J\bigg[\frac 1{ADE}\bigg]+J\bigg[\frac 1{BCE}\bigg]\right)\\
&=\frac 1{M^2(1+x-x_\gamma)}\left(J\bigg[\frac 1{ADE}\bigg]+J\bigg[\frac 1{ADE}\bigg]_{y\to-y}\right)\\
&=\left\{\frac 4{M^4(1+x-x_\gamma)[1+x-x_\gamma-y(1-x)]}\left(\frac 12\,J\bigg[\frac 1{AD}\bigg]+J\bigg[\frac 1{AE}\bigg]-J\bigg[\frac 1{CE}\bigg]_{y\to-y}\right)\right\}\\
&\hspace{4mm}+\{y\to-y\}
\end{split}
\end{align}


\newpage

\section{Computational methods}
\label{app:meth}

In this Appendix we show the approaches we used to evaluate the basic integrals listed in~Appendix \ref{app:Jbasic}.

\subsection{Feynman parametrization}
\label{app:FP}

With the help of the Feynman parametrization
\begin{equation}
\frac{1}{A_{1}^{\alpha_{1}}\cdots A_{n}^{\alpha_{n}}}=\frac{\Gamma(\alpha_{1}+\dots+\alpha_{n})}{\Gamma(\alpha_{1})\cdots\Gamma(\alpha_{n})}\int_{0}^{1}du_{1}\cdots\int_{0}^{1}du_{n}\frac{\delta(\sum_{k=1}^{n}u_{k}-1)u_{1}^{\alpha_{1}-1}\cdots u_{n}^{\alpha_{n}-1}}{\left[u_{1}A_{1}+\cdots+u_{n}A_{n}\right]^{\sum_{k=1}^{n}\alpha_{k}}}\,,
\end{equation}
we can prepare, for example, the following terms for further integration
\begin{align}
\begin{split}
\frac 1{AE}
&=\frac 2{E(2A)}
=2\int_0^1{\mathrm d}\alpha\,\frac 1{[(E-2A)\alpha+2A]^2}
\stackrel{(\ref{eq:EAB})}{=}2\int_0^1{\mathrm d}\alpha\,\frac 1{[2(A+\alpha B)+\alpha M^2x]^2}
\equiv 2\int_0^1\frac {{\mathrm d}\alpha}{\eta^2}
\end{split}\\
\frac 1{A^2E}
&=\frac 4{E(2A)(2A)}
=8\int_0^1{\mathrm d}\alpha\,\int_0^{1-\alpha}{\mathrm d}\beta\,\frac 1{[(E-2A)\alpha+2A]^3}
=8\int_0^1\frac {(1-\alpha){\mathrm d}\alpha}{\eta^3}\\
\frac 1{AE^2}
&=\frac 2{E^2(2A)}
=4\int_0^1\frac {\alpha\,{\mathrm d}\alpha}{\eta^3}\\
\frac 1{A^2E^2}
&=\frac 4{E^2(2A)^2}
=24\int_0^1\frac {\alpha(1-\alpha)\,{\mathrm d}\alpha}{\eta^4}\,,
\end{align}
where we have defined $\eta=2\,l\cdot(\alpha p+q)+\alpha M^2x+i\epsilon$\,.
Now, let us calculate the following (simplest) integral
\begin{equation}
\begin{split}
J\bigg[\frac 1{AE}\bigg]
&=2\,J\bigg[\int_0^1\frac {{\mathrm d}\alpha}{\eta^2}\bigg]
=\frac 2{4\pi}\int{\mathrm d}\Omega\int_0^1\frac {{\mathrm d}\alpha}{\eta^2}
=\frac 1{2\pi}\int_0^1{\mathrm d}\alpha\int\frac {{\mathrm d}\Omega}{\big[2\,l\cdot\underbrace{(\alpha p+q)}_u+\underbrace{{\alpha M^2x+i\epsilon}}_{2l_0 V}\big]^2}\\
&\hspace{-2cm}=\frac 1{4l_0^2}\int_0^1{\mathrm d}\alpha\int_{-1}^1\frac{{\mathrm d} z}{(u_0-|\vv{u}|z+V)^2}
=\frac 1{2l_0^2}\int_0^1\frac{{\mathrm d}\alpha}{u^2+V^2+2u_0V}
=\frac2{M^4}\int_0^1\frac{{\mathrm d}\alpha}{w_2\alpha^2+w_1\alpha+w_0}\,.
\end{split}
\end{equation}
We have introduced
\begin{align}
w_2&=x^2+\frac{4l_0 p_0x}{M^2}+\frac{4l_0^2m^2}{M^4}\stackrel{(\vv{r}=0)}{=}x^2+\frac{1}{2}x\left[(1-x)(1-y)-x_\gamma\right]+\frac 14{\nu^2x_\gamma}\\
w_1&=\frac{4l_0^2x}{M^2}+\frac{4l_0q_0 x}{M^2}-\frac{8 l_0^2m^2}{M^4}\stackrel{(\vv{r}=0)}{=}xx_\gamma+\frac{1}{2}x\left[(1-x)(1+y)-x_\gamma\right]-\frac 12{\nu^2x_\gamma}\\
w_0&=\frac{4l_0^2m^2}{M^4}\stackrel{(\vv{r}=0)}{=}\frac 14{\nu^2x_\gamma}\,.
\end{align}
The last integral can be evaluated as
\begin{equation}
\int_0^1\frac{{\mathrm d}\alpha}{w_2\alpha^2+w_1\alpha+w_0}
=\frac 1{\sqrt{w_1^2-4w_0w_2}}\log\left(\frac{2w_0+w_1+\sqrt{w_1^2-4w_0w_2}}{2w_0+w_1-\sqrt{w_1^2-4w_0w_2}}\right)
\equiv\frac{1}{\sqrt{\kappa}}\log\left(\frac{\rho+\sqrt{\kappa}}{\rho-\sqrt{\kappa}}\right)\,.
\end{equation}

The other integrals which belong to this section can be treated in the following way
\begin{align}
\frac 1{AB}
&=\frac {(-1)}{B(-A)}
=-\int_0^1{\mathrm d}\alpha\,\frac 1{[(A+B)\alpha-A]^2}\\
\frac 1{CE}
&=\frac {(-2)}{E(-2C)}
=-2\int_0^1{\mathrm d}\alpha\,\frac 1{[(E+2C)\alpha-2C]^2}
\stackrel{(\ref{eq:ECD})}{=}-2\int_0^1{\mathrm d}\alpha\,\frac 1{[2(C+\alpha D)+\alpha M^2(x_\gamma-1)]^2}\\
\frac 1{BC}
&=\frac {1}{[(A+C)-A]B}
=\int_0^1{\mathrm d}\alpha\,\frac 1{[(A+B)\alpha-B-\alpha\underbrace{(A+C)}_{(\ref{eq:AC})}]^2}\,.
\end{align}

\newpage

\subsection{Legendre polynomials and functions of the second kind}
\label{app:Legendre}

Finally, two basic integrals can be evaluated by expanding to the Legendre functions. We can write
\begin{equation}
\frac B A
=\frac{l\cdot p}{l\cdot q}
=\frac{l_0p_0-|\vv{l}||\vv{p}|\cos\theta_p}{l_0q_0-|\vv{l}||\vv{q}|\cos\theta_q}
=\frac{|\vv{p}|}{|\vv{q}|} \cdot \frac{\frac{p_0}{|\vv{p}|}-\cos\theta_p}{\frac{q_0}{|\vv{q}|}-\cos\theta_q}
\equiv\frac{|\vv{p}|}{|\vv{q}|} \cdot \frac{a-\cos\theta_p}{b-\cos\theta_q}\,,
\label{eq:BA}
\end{equation}
where we have introduced $a={p_0}/{|\vv{p}|}$ and $b={q_0}/{|\vv{q}|}$\,. Now, consider first two Legendre polynomials and Legendre functions of the second kind, i.e.\
\begin{equation}
P_0(x)=1\,,\;P_1(x)=x\,,
\end{equation}
\begin{equation}
Q_0(x)=\frac 12 \log\frac{x+1}{x-1}\,,\;Q_1(x)=xQ_0(x)-1\,.
\end{equation}
The numerator in (\ref{eq:BA}) can thus be rewritten in terms of the Legendre polynomials and the denominator can be expanded in the following way%
\footnote{We use the {\it hat} sign to stand for the unit vector, i.e.\ $\hat{\vv{l}}=\vv{l}/|\vv{l}|$\,.}
\begin{equation}
\frac{a-\cos\theta_p}{b-\cos\theta_q}
\equiv\frac{a-\hat{\vv{l}}\cdot\hat{\vv{p}}}{b-\hat{\vv{l}}\cdot\hat{\vv{q}}}
=\left[a P_0\Big(\hat{\vv{l}}\cdot\hat{\vv{p}}\Big)-P_1\Big(\hat{\vv{l}}\cdot\hat{\vv{p}}\Big)\right]\sum_m(2m+1)P_m\Big(\hat{\vv{l}}\cdot\hat{\vv{q}}\Big)Q_m(b)\,.
\label{eq:BAexpand}
\end{equation}
Since there is a useful integral formula for unit vectors $\vv {n}$ and $\vv {n}_i$
\begin{equation}
\int P_m(\vv{n}\cdot\vv{n}\hspace{-0.3mm}_1)P_{m^\prime}(\vv{n}\cdot\vv{n}\hspace{-0.3mm}_2)\,{\mathrm d}\Omega\left(\vv{n}\hspace{0.1mm}\right)
=\frac{4\pi}{(2m+1)}\delta_{mm^\prime}P_m(\vv{n}_1\cdot\vv{n}_2)\,,
\end{equation}
the infinite sum in (\ref{eq:BAexpand}) reduces in the final result to only two terms
\begin{equation}
\frac{|\vv{q}|}{|\vv{p}|}J\bigg[\frac BA\bigg]
=\frac 1{4\pi}\frac{|\vv{q}|}{|\vv{p}|}\int\frac BA \,{\mathrm d}\Omega
=a Q_0(b)P_0\Big(\hat{\vv{p}}\cdot\hat{\vv{q}}\Big)-Q_1(b)P_1\Big(\hat{\vv{p}}\cdot\hat{\vv{q}}\Big)
=a Q_0(b)-\left(\hat{\vv{p}}\cdot\hat{\vv{q}}\right)Q_1(b)\,.
\end{equation}
We can differentiate the previous terms in order to get the last missing piece, since
\begin{equation}
\frac B {A^2}
=\frac{l\cdot p}{(l\cdot q)^2}
\equiv\frac{|\vv{p}|}{l_0|\vv{q}|^2} \cdot \frac{a-\cos\theta_p}{(b-\cos\theta_q)^2}
=-\frac{1}{l_0|\vv{q}|}\cdot\frac{\partial}{\partial b}\left[\frac{|\vv{p}|}{|\vv{q}|} \cdot \frac{a-\cos\theta_p}{b-\cos\theta_q}\right]
=-\frac{1}{l_0|\vv{q}|}\frac{\partial}{\partial b}\left(\frac BA\right)\,.
\end{equation}
Hence
\begin{equation}
J\bigg[\frac B {A^2}\bigg]
=-\frac{1}{l_0|\vv{q}|}\frac{\partial}{\partial b}\left(J\bigg[\frac BA\bigg]\right)\,.
\end{equation}
The results can be written in the form
\begin{align}
J\bigg[\frac BA\bigg]
&=\frac{p_0}{|\vv{q}|}\,Q_0\bigg(\frac{q_0}{|\vv{q}|}\bigg)
+\frac{\frac 12(M^2x-2m^2)-p_0q_0}{|\vv{q}|^2}\,Q_1\bigg(\frac{q_0}{|\vv{q}|}\bigg)\\
J\bigg[\frac B {A^2}\bigg]
&=-\frac{p_0}{l_0|\vv{q}|^2}\,Q_0^\prime\bigg(\frac{q_0}{|\vv{q}|}\bigg)
-\frac{\frac 12(M^2x-2m^2)-p_0q_0}{l_0|\vv{q}|^3}\,Q_1^\prime\bigg(\frac{q_0}{|\vv{q}|}\bigg)\,,
\end{align}
where
\begin{equation}
Q_0^\prime(x)
=\frac{1}{1-x^2}\,,\quad
Q_1^\prime(x)
=Q_0(x)+xQ_0^\prime(x)\,.
\end{equation}


\newpage

\section{Basic \texorpdfstring{$J$}{J} terms}
\label{app:Jbasic}

\twocolumngrid

In this Appendix we list the results of the basic set of integrals generated by acting of the operator $J$ on the desired combinations of variables A, $\dots$, F in terms of (\ref{eq:Jint}). First, we define a useful logarithmic function
\begin{equation}
\begin{split}
L(a,b)
&\equiv\frac 1{\sqrt{a^2-b}}\log\left|\frac{a+\sqrt{a^2-b}}{a-\sqrt{a^2-b}}\right|\\
&\hspace{-2.3mm}\stackrel{a>0}{=}\frac 1{\sqrt{a^2-b}}\log\left|-1+\frac 2{1-\sqrt{1-\frac b{a^2}}}\right|
\end{split}
\end{equation}
and variables, in which the results have the simple forms
\begin{align}
v_1&=\frac{1}4\left[(1-x)(1+y)-x_\gamma\right]\\
v_2&=\frac{1}4\left[(1-x)(1-y)-x_\gamma\right]\\
v_0&=v_1+v_2+x=\frac 12 (1+x-x_\gamma)\\\notag\\
\rho&=xx_\gamma+2xv_1\\
\rho^\prime&=xx_\gamma-2(1-x_\gamma)v_1\\
\kappa&=\rho^2-\nu^2xx_\gamma\\
\omega&=-v_0^2+x+\frac 14 (1-x)^2y^2\\\notag\\
\xi_0&=\nu^2(v_0-1)+\rho\\
\xi_1&=\rho(v_0-1)+xx_\gamma\\
\xi_2
&=\frac{\kappa}{2x}-\frac {x_\gamma}2\xi_0\\
\xi&=1-\frac {12}{\nu^2x_\gamma}\left(v_1^2-\frac{\xi_2^2}{\kappa}\right)\,.
\end{align}

Using standard integration techniques we find the following integrals
\begin{align}
J[1]
&=1\\
J\bigg[\frac 1A\bigg]
&=\frac 2{M^2}L(2v_1,\nu^2x_\gamma)\\
J\bigg[\frac 1{A^2}\bigg]
&=\frac {16}{M^4\nu^2x_\gamma}\\
J\bigg[\frac 1E\bigg]
&=\frac 1{2M^2}L(v_0,x)\\
J\bigg[\frac 1{E^2}\bigg]
&=\frac 1{M^4x}\,.
\end{align}

With the help of the Feynman parametrization (for details see Appendix \ref{app:FP}) we are able to calculate
\begin{align}
J\bigg[\frac 1{AB}\bigg]
&=\frac 8{M^4x_\gamma}L(x,\nu^2x)\\
J\bigg[\frac 1{AE}\bigg]
&=\frac 2{M^4}L(\rho,\nu^2xx_\gamma)\\
J\bigg[\frac 1{CE}\bigg]
&=\frac 2{M^4}L(\rho^\prime,\nu^2xx_\gamma)\\
J\bigg[\frac 1{BC}\bigg]
&=\frac 8{M^4}L(\omega,\nu^2x_\gamma \omega)\\
J\bigg[\frac 1{AE^2}\bigg]
&=\frac {4\xi_1}{M^6\kappa x}+\frac {2\xi_2}{M^2\kappa}\,J\bigg[\frac 1{AE}\bigg]\\
J\bigg[\frac 1{A^2E}\bigg]
&=\frac {32\xi_2}{M^6\nu^2\kappa x_\gamma}+\frac {4\xi_1}{M^2\kappa}\,J\bigg[\frac 1{AE}\bigg]\label{eq:A2E}
\end{align}
as well as the integral, which did not need to be evaluated in the original work~\cite{Mikaelian:1972yg} due to the systematic neglecting of the terms of order higher than~$m^2$
\begin{equation}
\begin{split}
J\bigg[\frac 1{A^2E^2}\bigg]
&=\frac{16}{M^8\kappa}\left(\frac{4v_1^2}{\nu^2x_\gamma}+\frac{v_0^2}x+\xi\right)\\
&-\frac{4}{M^4\kappa}(4v_0v_1+\rho \xi)\,J\bigg[\frac 1{AE}\bigg]\,.
\end{split}
\label{eq:A2E2}
\end{equation}

Using the expansion to the Legendre polynomials and functions of the second kind (see Appendix \ref{app:Legendre}), we find
\begin{align}
J\bigg[\frac BA\bigg]
&=-\frac{\zeta_1}{\zeta}
+\frac {M^2x_\gamma\zeta_2}{2\zeta} J\bigg[\frac 1A\bigg]\\
J\bigg[\frac B{A^2}\bigg]
&=\frac {8\zeta_2}{M^2{\nu^2}\zeta}
-\frac{\zeta_1}{\zeta} J\bigg[\frac 1A\bigg]\,,
\end{align}
where we have introduced
\begin{align}
\zeta_1&=2xx_\gamma-\nu^2x_\gamma-4v_1v_2\\
\zeta_2&=2xv_1-\nu^2(v_0-x)\\
\zeta&=4v_1^2-\nu^2x_\gamma\,.
\end{align}

\newpage
\onecolumngrid

We can extract the divergent parts of the integrals (\ref{eq:A2E}) and (\ref{eq:A2E2}) through
\begin{align}
\begin{split}
J\bigg[\frac 1{A^2E}\bigg]
&=\frac 1{M^2x}\,J\bigg[\frac 1{A^2}\bigg]
-\frac {16\xi_0}{M^6\nu^2\kappa}+\frac {4\xi_1}{M^2\kappa}\,J\bigg[\frac 1{AE}\bigg]
\end{split}\\
\begin{split}
J\bigg[\frac 1{A^2E^2}\bigg]
&=\frac 1{M^4x^2}\,J\bigg[\frac 1{A^2}\bigg]+\frac{16}{M^8\kappa}\bigg[1+\frac1x
\left(v_0^2-2-\frac{6\xi_0}{\nu^2}\right)+\frac 2{\nu^2}\left(4v_1+x_\gamma+\frac{3x_\gamma\xi_0^2}{2\kappa}\right)\bigg]\\
&-\frac{4}{M^4\kappa}(4v_0v_1+\rho \xi)\,J\bigg[\frac 1{AE}\bigg]\,.
\end{split}
\end{align}
The above formulas have a very convenient form and are to be substituted into (\ref{eq:TrC}).

The divergent integrals alone have to be integrated over $x_\gamma$ analytically. The calculation is done in detail in Ref.~\cite{Mikaelian:1972yg} and the results can be written in a simple form
\begin{align}
\int J\bigg[\frac 1{A^2}\bigg]{\mathrm d}x_\gamma
&=\frac{16}{\nu^2M^4}\bigg[\log\frac m \lambda+\log\frac{2x_\gamma^\text{max}}{(1-x)(1+y)}\bigg]
\label{eq:JA2div}\\
\begin{split}
\int J\bigg[\frac 1{AB}\bigg]{\mathrm d}x_\gamma
&=-\frac{8}{M^4x\beta}\bigg[\log\frac{m}{\lambda}+\log\frac{2x_\gamma^\text{max}}{(1-x)}\bigg]\log(\gamma)+\frac{4}{M^4x\beta}K(x,y)\,,
\end{split}
\label{eq:JABdiv}
\end{align}
where
\begin{equation}
K(x,y)
=\bigg[2\log\bigg(\frac{y+\beta}{2\beta}\bigg)+\log\frac{\nu^2}x\bigg]\log(\gamma)
-\text{Li}_2\bigg[\frac{\gamma(y-\beta)}{y+\beta}\bigg]+\text{Li}_2\bigg[\frac{y-\beta}{\gamma(y+\beta)}\bigg]\,.
\label{eq:Kxy}
\end{equation}
These terms are to be used to evaluate $\int J[\text{Tr}_\text{D}]\,{\mathrm d}x_\gamma$\,.

\input{paper.bbl}

\end{document}

%% file: paper.bbl
%